\documentclass[aip,apl
amsmath,amssymb,reprint]{revtex4-1}

\usepackage{graphicx}
\usepackage{caption}
\usepackage{bm}
\usepackage[version=4]{mhchem}
\usepackage{xr}
\draft 

\begin{document}

\title{Fabrication of Metal Air Bridges for Superconducting Circuits using Two-photon Lithography}

\author{Yi-Hsiang Huang}
\affiliation{Department of Electrical and Computer Engineering, University of Maryland, College Park, Maryland 20742, USA}
\affiliation{Laboratory for Physical Sciences, 8050 Greenmead Drive, College Park, Maryland 20740, USA}
\affiliation{Quantum Materials Center, University of Maryland, College Park, Maryland 20742, USA}

\author{Haozhi Wang}
\affiliation{Laboratory for Physical Sciences, 8050 Greenmead Drive, College Park, Maryland 20740, USA}
\affiliation{Quantum Materials Center, University of Maryland, College Park, Maryland 20742, USA}
\affiliation{Department of Physics, University of Maryland, College Park, Maryland 20742, USA}

\author{Zhuo Shen}
\affiliation{Laboratory for Physical Sciences, 8050 Greenmead Drive, College Park, Maryland 20740, USA}
\affiliation{Quantum Materials Center, University of Maryland, College Park, Maryland 20742, USA}
\affiliation{Department of Physics, University of Maryland, College Park, Maryland 20742, USA}

\author{Austin Thomas}
\altaffiliation{Current address: Booz Allen Hamilton, 8283 Greensboro Drive, Mclean, Virginia, 22102.}
\affiliation{Laboratory for Physical Sciences, 8050 Greenmead Drive, College Park, Maryland 20740, USA}

\author{C. J. K. Richardson}
\affiliation{Laboratory for Physical Sciences, 8050 Greenmead Drive, College Park, Maryland 20740, USA}
\affiliation{Department of Materials Science and Engineering, University of Maryland, College Park, Maryland 20742, USA}

\author{B. S. Palmer}
\email[Author to whom correspondence should be addressed: ]{bpalmer@umd.edu}
\affiliation{Laboratory for Physical Sciences, 8050 Greenmead Drive, College Park, Maryland 20740, USA}
\affiliation{Quantum Materials Center, University of Maryland, College Park, Maryland 20742, USA}
\affiliation{Department of Physics, University of Maryland, College Park, Maryland 20742, USA}


\begin{abstract}
Extraneous high frequency chip modes parasitic to superconducting quantum circuits can result in decoherence when these modes are excited. To suppress these modes, superconducting air bridges (AB) are commonly used to electrically connect ground planes together when interrupted by transmission lines. Here, we demonstrate the use of two-photon photolithography to build a supporting 3D resist structure in conjunction with a lift-off process to create AB. The resulting  aluminum AB, have a superconducting transition temperature $T_{c} = 1.08$ K and exhibit good mechanical strength up to lengths of 100 $\mu$m. A measurable amount of microwave loss is observed when 35 AB were placed over a high-$Q$ Ta quarter-wave coplanar waveguide resonator.
\end{abstract}


\maketitle 

Micro-fabricated superconducting circuits have emerged as a leading candidate for quantum computers. \cite{2019Arute,2023Kim} However, designing a chip with increased number of qubits requires careful consideration of signal routing and elimination of unwanted modes. Coplanar waveguides (CPW), which are used for transmission lines and resonators,\cite{2003Day, 2005Pozar} are one of the constituent parts of a planar superconducting  circuit. When transmitting signals through a CPW, parasitic slotline modes can arise due to uneven voltages induced on either side of the ground planes separated by the center metal trace.\cite{2005Ponchak} To suppress excitation of slotline modes, wire-bonds are one commonly used technique to electrically connect the ground planes together. Despite their easy installation, wire-bonds are proven to be ineffective shunts due to their millimeter lengths.\cite{2011Wenner} Instead, superconducting free-standing crossovers, or air bridges (AB), are more effective in suppressing the parasitic modes as well as providing a way of signal routing.\cite{2014Chen,2025Bu} \par
Various fabrication methods for AB have been explored and reported in the literature.\cite{2006Girgis,2012Lankwarden,2013Abuwasib,2014Chen,2018Dunsworth,2021Jin,2022Janzen,2022Sun,2023Stavenga,2025Alegria,2025Bu} Typical methods for fabrication of AB often include multiple layers of resist, thermal reflow, and etching of excess metals, adding complexity and inconsistencies to the fabrication process. Furthermore, the dimensions and the mechanical strength of the bridge structures are often limited by the height of the resist, which is determined by the spin and reflow steps in the fabrication process.\par
\begin{figure*}[htbp]
    \centering
    \includegraphics[width=.8\linewidth]{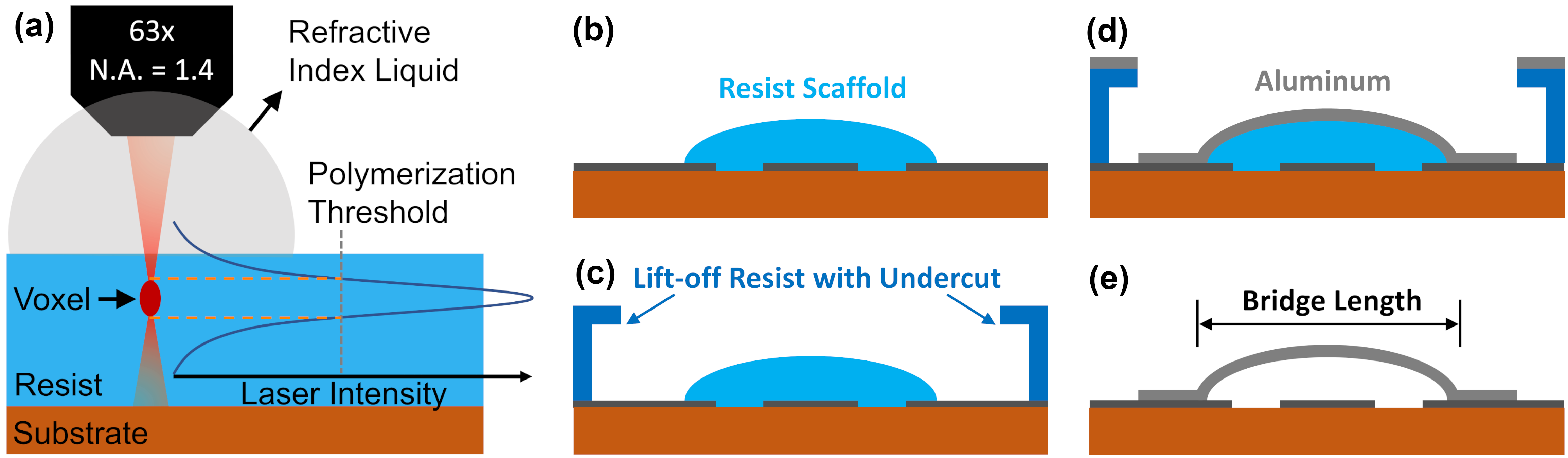}
    \caption{ (a) Illustration of the two-photon lithography process. A  large numerical aperture (N.A. = 1.4) objective lens focuses the $\lambda = 780$ nm laser light to a small volume (voxel), exceeding an intensity threshold and polymerizing the resist. (b)-(e) Our simplified fabrication process of AB: (b) Definition of the sacrificial scaffold following exposure by TPL and resist development. (c) Lift-off resist layer, defining the placement and dimensions of the bridge, is spun, exposed and developed. (d) Deposition of 400 - 550 nm of Al for the metal bridge. (e) Release of the Al bridge by stripping both layers of resist. \label{Fig1}}
\end{figure*}

Techniques towards building a 3D AB by using grayscale lithography or dielectric materials to form the supporting structure have been demonstrated in previous works,\cite{2006Girgis,2018Dunsworth,2022Janzen,2022Sun,2023Stavenga} but a thermal reflow of the resist is constantly required to achieve a smooth surface. Here, we present a new method that exploits two-photon lithography (TPL) to fabricate aluminum air bridges (Al AB). TPL is a direct laser writing (DLW) method that utilizes non-linear two-photon processes of the resist.\cite{1997Maruo} Fig. \ref{Fig1}(a) illustrates the basic working principle of TPL. In a two-photon process, the extent of polymerization of the underlying resist  depends on the laser intensity non-linearly. \cite{1931Goppert‐Mayer,1961Kaiser} As a result, resists only polymerize in a confined volume near the focal point (FP) of the laser (voxel) where the light intensity exceeds the polymerization threshold. By moving the FP through the resist, three-dimensional lithography is performed. TPL has found applications in various fields, including micro-electromechanical systems, photonics, and biomedical engineering.\cite{2020Lei,2020vanderVelden,2021Harinarayana} With its ability to create sub-micron scale 3D structures,\cite{2013Cao,2016Guney,2020vanderVelden} TPL potentially enables superconducting circuits with complexity not previously achievable using traditional two-dimensional fabrication techniques. \par

We start by presenting the key concepts of our fabrication process for Al AB in Fig. \ref{Fig1}(b)-(e). First, a layer of negative-tone photoresist (AZ 15nXT) with a nominal thickness of 5.5 $\mu$m was spun and baked on a device with patterned metal. The sacrificial scaffolds initially supporting the bridge structures were defined by exposing the resist using TPL (Photonic Professional GT2, Nanoscribe). The laser used, had a mean power of 50 mW, a wavelength of 780 nm, and illuminated the resist through an objective lens with a magnification of 63x and a numerical aperture of N.A. = 1.4. To obtain this N.A., the region between the objective and sample consisted of a matching refractive index liquid (Cargille) with $n_D$ = 1.6260 $\pm$ 0.0002. After exposing and developing the solid resist  scaffolds, a subsequent lift-off layer of 7 $\mu$m photoresist (AZ nLOF 2070) was spun, baked, and exposed using a direct-write laser lithography tool (MLA150, Heidelberg). After developing this top layer of resist, which creates the aperture for the metal AB, the device was mounted  in an electron beam evaporator (MEB550S, Plassys) for deposition of the metal AB. Prior to the deposition, an \textit{in situ} aggressive argon ion mill was applied to remove the native oxide layer of the base metal and ensure good electrical contact. Then, 400 - 550 nm of Al was deposited at a rate of 0.5 nm/s to form the metal bridges. To remove the cross-linked resist formed during the argon ion mill step and assist the  lift-off process, a  reactive ion etch process using oxygen plasma was performed. \cite{1978Okuyama,2014Chen} Finally, the two resists were dissolved and the AB were released by immersing the device in two subsequent baths of photoresist stripper (NI555, TechniStrip) at 80$^{\circ}$C. \par

\begin{figure}[htbp]
    \centering
    \includegraphics[width=.8\linewidth]{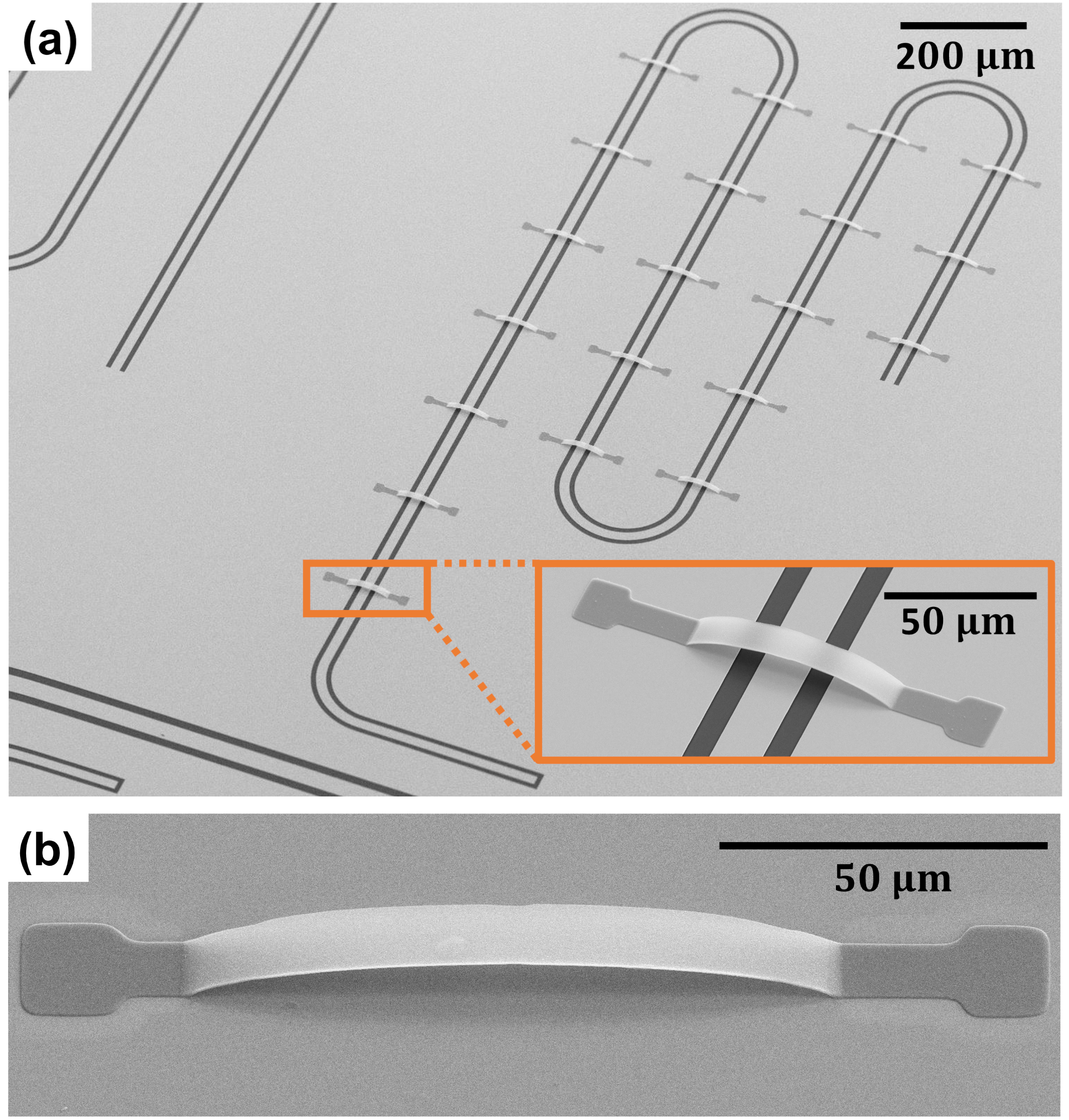}
    \caption{SEM micrographs of TPL-fabricated Al AB. (a) Twenty free-standing Al crossovers connecting the ground plane of a Ta quarter-wave coplanar waveguide resonator. Inset shows a zoom-in of a single bridge in (a). (b) A 100 $\mu$m long Al bridge fabricated on Si. \label{Fig2}}
\end{figure}

Using this TPL process, Al AB have been fabricated on both patterned reflective metal films on an optically transparent substrate such as sapphire and unpatterned substrates such as Si. Fig. \ref{Fig2} (a) shows a SEM micrograph of 20 AB, each AB has a length of 70 $\mu$m and a width of 12 $\mu$m creating a bridge for electrical ground signals, across a 6.15 GHz Ta quarter-wave CPW resonator. AB with lengths up to 100 $\mu$m have been reliably fabricated  (see Fig. \ref{Fig2} (b)). \par

To initially characterize the electrical properties of the resulting Al AB, 29 bridges in series were fabricated on a sapphire substrate and the resistance was measured using a four-probe measurement technique. The room temperature resistivity for the deposited Al AB was estimated to be $\rho = 4.4\ \mathrm{\mu \Omega \cdot cm}$, though we note that this estimate does not take into account the exact curved shape of the AB. The  device was then cooled in a cryogen-free dilution refrigerator; a residual resistivity ratio of 3.85 and a sharp superconducting transition temperature of $T_{c} = 1.08$ K were measured for the series of 29 Al AB. \par

The microwave loss, pertaining to the addition of our Al AB, was then characterized by fabricating three different numbers ($N_{AB}$) of Al AB  over Ta quarter-wave CPW resonators. For this work, a 200 nm thick tantalum film was deposited in a molecular-beam epitaxy system on a sapphire substrate. Four resonators coupled to a common transmission line were subsequently defined by photolithography and the Ta was etched  using \ce{CF4}/\ce{CHF3} in an inductively coupled plasma etcher. The CPW resonators nominally had a center trace of 16 $\mu$m, gaps of 8 $\mu$m, and different lengths such that the resonance frequencies ($f_{r}$) ranged from 5.38 to 6.15 GHz. We then fabricated 8, 20, and 35 AB, respectively, across three of the four resonators (see Fig. \ref{FigS2}(a) in the supplementary material for an optical image of the device).\cite{fnote1} To measure the low-temperature microwave loss, the chip was then packaged and cooled\cite{fnote1,2023Huang} in a cryogen-free dilution refrigerator with a base temperature of 10 mK. Using a vector network analyzer, the in-phase and out-of-phase S$_{21}= V_{out}/V_{in}$ was measured as a function of frequency at different powers and temperatures. The internal quality factors, $Q_i$, were extracted by fitting S$_{21}$ using the diameter correction method.\cite{fnote1,2012Khalil} \par

\begin{figure}[htbp]
    \centering
    \includegraphics[width=.85\linewidth]{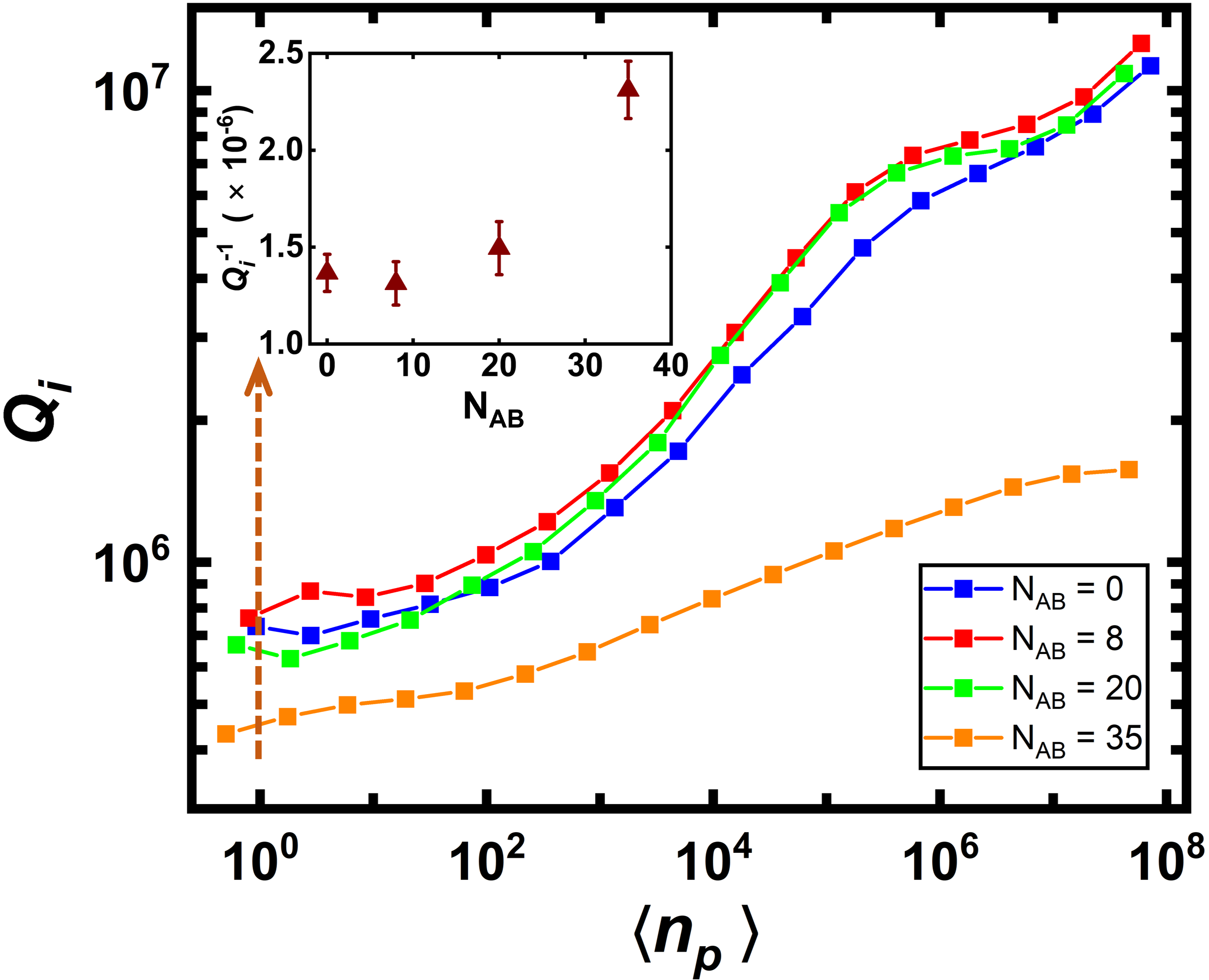}
    \caption{Extracted internal quality factor $Q_i$ as a function of average photon number $\langle n_p \rangle$ for thin film Ta resonators with $N_{AB}$ = 0, 8, 20, 35. Inset shows $Q_i^{-1}$ as a function of $N_{AB}$ at single photon levels. \label{Fig3}}
\end{figure}

Fig. \ref{Fig3} shows the extracted $Q_i$ as a function of the average stored photon number $\langle n_p \rangle$ for the four different resonators at $T=10$ mK. At low powers corresponding to $\langle n_p \rangle\simeq 1$,  the three resonators with $N_{AB}$ = 0, 8, 20 had $Q_{i} \sim 7.5\times 10^5$. These three resonators display a similar and weak increase of $Q_{i}$ with increasing $\langle n_p \rangle$, reaching $Q_{i}\sim 10^7$ at stored powers corresponding to $\langle n_p \rangle\sim 10^8$. On the other hand, the resonator device with $N_{AB}$ = 35, displays more loss with $Q_i \simeq 4.5 \times 10^5$ at $\langle n_p \rangle\simeq 1$ and $Q_i \simeq 1.5 \times 10^6$ at large powers. This observed loss for $N_{AB} = 35$ at $\langle n_p \rangle\sim 1$ is similar to the loss previously reported by other AB fabrication techniques (see inset Fig.~\ref{Fig3}). \cite{2014Chen,2018Dunsworth,2022Janzen} We note the dependence of $Q_i$ on $\langle n_p \rangle$ corresponding to the $N_{AB}$ = 35 resonator can be well modeled as a power dependent loss similar to the other three resonators plus a power independent loss, $Q_{const}^{-1} \simeq 5\times 10^{-7}$. 

\begin{figure}[htbp]
    \centering
    \includegraphics[width=.85\linewidth]{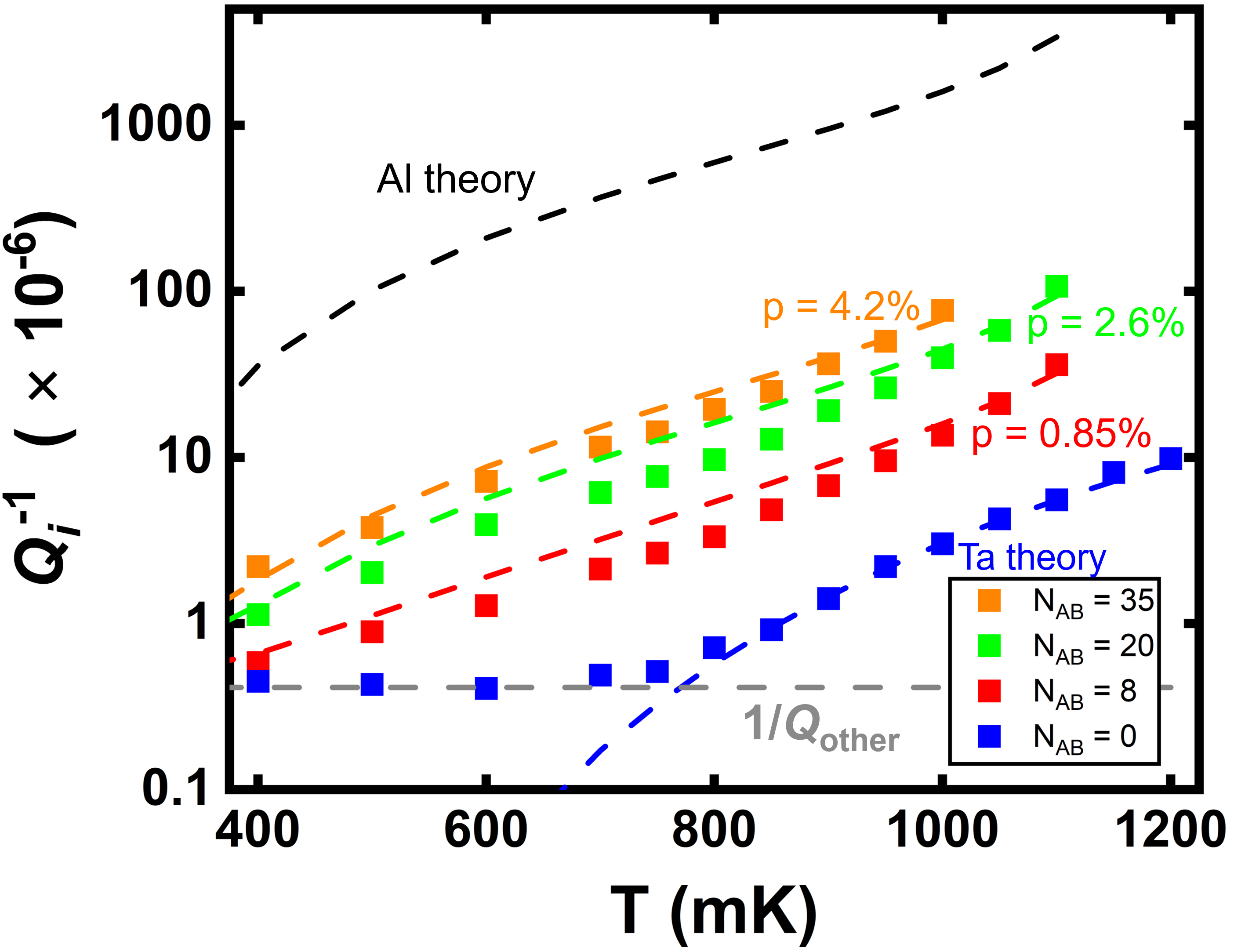}
    \caption{$Q_i^{-1}$ as a function of temperature from $T= 400$ mK to $T= 1.2$ K and with a resonator drive strength $\langle n_p \rangle\simeq 1$. Dashed lines represent theoretical calculation for Ta (blue), Al (black), and a temperature independent loss $Q_{other}^{-1}$ (gray). We extract the percentage ($p$) of inductive energy stored in AB using Eq. \ref{eq:p}. \label{Fig4}}
\end{figure}

To determine if this source of loss $Q_{const}^{-1}$ is from inductive losses, we have determined the percentage of inductive energy stored in the AB for each $N_{AB}$. Since there is a significant   difference in the superconducting gaps ($\Delta$) between Al and Ta, the degradation of quality factors at elevated temperatures above 400 mK, depends on how much inductive energy is stored in the Al AB. Fig. \ref{Fig4} shows measurements of the loss ($Q_i^{-1}$) as a function of temperature from $T = 400$ mK to above 1 K at $\langle n_p \rangle\simeq 1$. The $Q_i^{-1}$ exhibits a clear dependence on $N_{AB}$: with more air bridges added, the resonators display more loss at high temperatures. We model this dependence of $Q_i^{-1}$ on  $N_{AB}$ by assuming a percentage ($p$) of inductive energy stored in the Al AB and calculate the microwave loss of Al and Ta, respectively, as a function of temperature. 

Using the theory of Mattis and Bardeen, \cite{1958Mattis,1977Broom} $Q^{-1}$ associated with quasiparticles for a superconducting resonator with film thickness $d$ are given by
\begin{equation}
    Q^{-1} = \frac{\alpha}{2}\frac{\sigma_1}{\sigma_2}\left( 1+\frac{2d/\lambda}{\sinh(2d/\lambda)}\right)
    \label{eq:Q_MB}
\end{equation}
where $\alpha = L_k / (L_k + L_g)$ is the fraction of kinetic to total (kinetic + geometric) inductance. The real and imaginary parts of the complex conductivity $\sigma = \sigma_1 - i\sigma_2$, and the London penetration depth $\lambda$ as a function of $T$ are given by
\begin{widetext}
\begin{subequations}
    \begin{align}
    \frac{\sigma_1(T)}{\sigma_n} &= \frac{2}{hf_r}\int_{\Delta}^{\infty}\left(f(\epsilon)-f(\epsilon+hf_r)\right) \left(1+\frac{\Delta^2}{\epsilon(\epsilon+hf_r)}\right)\frac{\epsilon}{\sqrt{\epsilon^2-\Delta^2}}\frac{\epsilon+hf_r}{\sqrt{(\epsilon+hf_r)^2-\Delta^2}} \, d\epsilon \\
    \frac{\sigma_2(T)}{\sigma_n} &= \frac{1}{hf_r}\int_{\Delta-hf_r}^{\Delta}\left(1-2f(\epsilon+hf_r)\right) \frac{\epsilon(\epsilon + hf_r)+\Delta^2}{\sqrt{\Delta^2-\epsilon^2}\sqrt{(\epsilon+hf_r)^2-\Delta^2}}\, d\epsilon \\
    \lambda(T) &= \frac{1}{\sqrt{2\pi \mu_0f_r\sigma_2(T)}}.
    \end{align}
    \label{eq:MB}
\end{subequations}
\end{widetext}
Here, $f(\epsilon)$ is the Fermi-Dirac distribution and $\sigma_n$ is the normal state conductivity of the superconductor under consideration. By using $\Delta_{Al} = 182 \, \mathrm{\mu eV}$, $\Delta_{Ta} = 607 \, \mathrm{\mu eV}$, $\sigma_n^{Al} = 3.8 \times 10^7 \, \Omega^{-1} \cdot \mathrm{m}^{-1}$, $\sigma_n^{Ta} = 7.7 \times 10^6 \, \Omega^{-1} \cdot \mathrm{m}^{-1}$, $\alpha_{Al} =$ 0.022, and $\alpha_{Ta} =$ 0.018, we theoretically calculate $Q^{-1}$ as a function of $T$ (see Fig.\ref{Fig4}) for Al (black dashed line) and Ta (blue dashed line). The percentage ($p$) of inductive energy stored in the AB is then extracted for each $N_{AB}$ data set  using
\begin{equation}
    Q_i^{-1} = p \, Q_{Al \, theory}^{-1} + (1-p) \, Q_{Ta \, theory}^{-1} + Q_{other}^{-1}
    \label{eq:p}
\end{equation}
where $Q_{other}^{-1}$ is an unknown temperature independent loss. The fitted $p$'s range from 0.85\% for $N_{AB}=8$ up to 4.2 \% for $N_{AB} =35$ (see dashed lines in Fig.\ref{Fig4}). A similar but slightly reduced $p$'s were found when performing electromagnetic simulations using Ansys' high frequency simulation software (HFSS).\cite{fnote1} 

By using these $p$ values, we can determine if the $Q_{i}$'s measured at 10 mK are consistent with inductive losses. If we assume the power independent loss ($Q_{const}^{-1}$) observed in the $N_{AB} = 35$ resonator was inductive in nature and present in the $N_{AB}= 8$ resonator, the expected $Q_{i}$ is \[Q_{i}=\left(\frac{p_{N_{AB}\,=\,8}}{p_{N_{AB}\,=\,35}} Q_{const}^{-1}+10^{-7}\right)^{-1}=5\times10^{6}\] for $\langle n_p \rangle=10^{8}$, a value that is half of what is observed. This suggests that the observed loss with $N_{AB}=35$ is not inductive in nature. Further tests including adjusting the placement of the AB and a larger sampling of $N_{AB}$ would be needed to discern the loss mechanism.



In conclusion, we have used TPL and a lift-off process to create Al air bridges on superconducting CPW resonators, demonstrating a new technique for making AB. Our method can potentially be extended to create superconducting circuits with increased complexity enabled by the 3D capability of TPL. We confirmed the fabricated Al bridges to be superconducting and measured the loss when AB were placed over resonators. We have observed that with $N_{AB} = 35$ a measurable amount of loss is observed at 10 mK. Currently  the loss mechanism is not consistent with being inductive in nature though further studies need to be conducted to fully understand the mechanism. Finally, the  exposed TPL resist resolution  perpendicular to the substrate is limited by the resist voxel size of 3 $\mu$m.\cite{fnote1} This limits our ability in creating an integrated lithographic process where the aperture at the top of the resist is defined  in conjunction with the scaffold using TPL. Such an integrated process would be desired to further simplify the fabrication process.\par

\begin{acknowledgments}
The authors thank Karen Grutter, Ryan Sochol, and Trisha Chakraborty for insightful discussions and suggestions.
\end{acknowledgments}

\section*{Data Availability}
The data that support the findings of this study are available from the corresponding author upon reasonable request.

\bibliography{ref}

\end{document}


\title{Supplementary Material for "Fabrication of Metal Air Bridges for Superconducting Circuits using Two-photon Lithography"}

\author{Yi-Hsiang Huang}
\affiliation{Department of Electrical and Computer Engineering, University of Maryland, College Park, Maryland 20742, USA}
\affiliation{Laboratory for Physical Sciences, 8050 Greenmead Drive, College Park, Maryland 20740, USA}
\affiliation{Quantum Materials Center, University of Maryland, College Park, Maryland 20742, USA}

\author{Haozhi Wang}
\affiliation{Laboratory for Physical Sciences, 8050 Greenmead Drive, College Park, Maryland 20740, USA}
\affiliation{Quantum Materials Center, University of Maryland, College Park, Maryland 20742, USA}
\affiliation{Department of Physics, University of Maryland, College Park, Maryland 20742, USA}

\author{Zhuo Shen}
\affiliation{Laboratory for Physical Sciences, 8050 Greenmead Drive, College Park, Maryland 20740, USA}
\affiliation{Quantum Materials Center, University of Maryland, College Park, Maryland 20742, USA}
\affiliation{Department of Physics, University of Maryland, College Park, Maryland 20742, USA}

\author{Austin Thomas}
\affiliation{Laboratory for Physical Sciences, 8050 Greenmead Drive, College Park, Maryland 20740, USA}

\author{C. J. K. Richardson}
\affiliation{Laboratory for Physical Sciences, 8050 Greenmead Drive, College Park, Maryland 20740, USA}
\affiliation{Department of Materials Science and Engineering, University of Maryland, College Park, Maryland 20742, USA}

\author{B. S. Palmer}
\affiliation{Laboratory for Physical Sciences, 8050 Greenmead Drive, College Park, Maryland 20740, USA}
\affiliation{Quantum Materials Center, University of Maryland, College Park, Maryland 20742, USA}
\affiliation{Department of Physics, University of Maryland, College Park, Maryland 20742, USA}

\begin{abstract}
In the supplementary material, we detail the measurement setup and the analysis method used to extract the internal quality factors, $Q_i$. The voxel size of the resist used to form the bridge scaffolds was estimated. Knowledge of the resist voxel size can benefit the bridge designing process. To verify measurements of the percentage of inductive energy stored in the air bridges,  electromagnetic simulations for the resonator/air bridge device were performed and  discussed here. 
\end{abstract}

\maketitle

\section{Microwave Loss Measurement setup}
\begin{figure*}
    \centering
    \includegraphics[width=.5\linewidth]{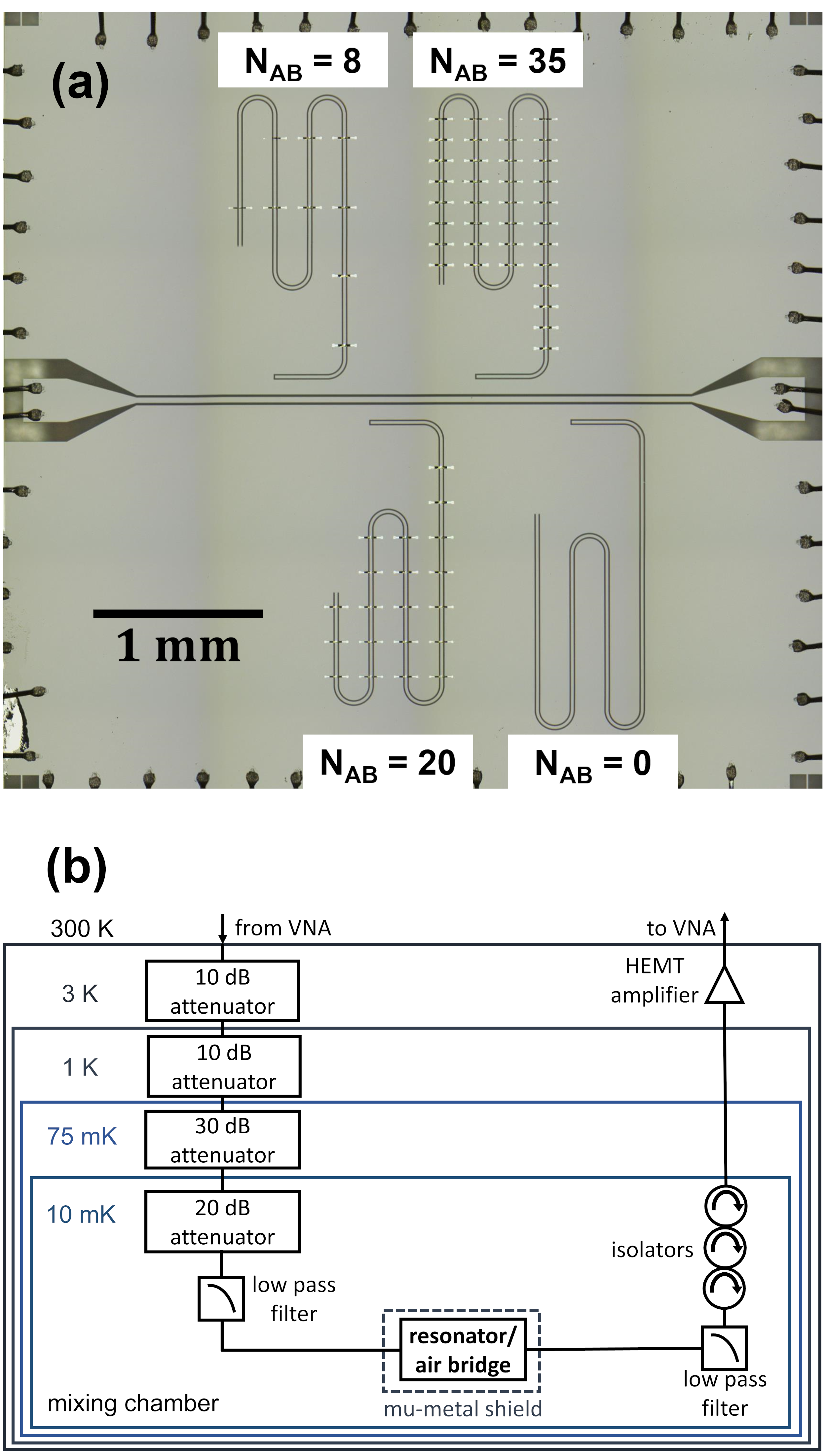}
    \caption{(a) Stitched optical image of the Ta resonator/Al AB device. (b) Measurement setup. Multiple attenuators at different temperature stages allow the resonators to achieve $\langle n_p \rangle \sim 1$. \label{FigS2}}
\end{figure*}

The resonator/air bridge device (Fig. \ref{FigS2}(a)) was mounted to the mixing chamber stage of a cryogen-free dilution refrigerator (LD400, Bluefors) and cooled to a base temperature of 10 mK. Fig. \ref{FigS2}(b) shows the dilution refrigerator and the measurement setup. A vector network analyzer (VNA) was used to measure the transmission coefficient $S_{21}$ at different input powers. To reduce Johnson-Nyquist noise from higher temperatures, attenuators at different temperature stages, for a total attenuation of 70 dB, were mounted on the input microwave cable going to the device. On the output microwave line,  three isolators on the mixing chamber were used before the signals were amplified with a high-electron-mobility transistor (HEMT) amplifier at the 3K stage. At room temperature the signal was further amplified with a low-noise amplifier  before going into port 2 of the VNA. To shield the device from blackbody radiation, cans made of Al or Au plated Cu were bolted to every stage except the 75 mK cold plate heat exchanger plate. To suppress stray magnetic fields, an open ended cylinder of Amumetal 4K (Amuneal) surrounded the device on the mixing chamber.

\section{Data Analysis}
\begin{figure*}
    \centering
    \includegraphics[width=.5\linewidth]{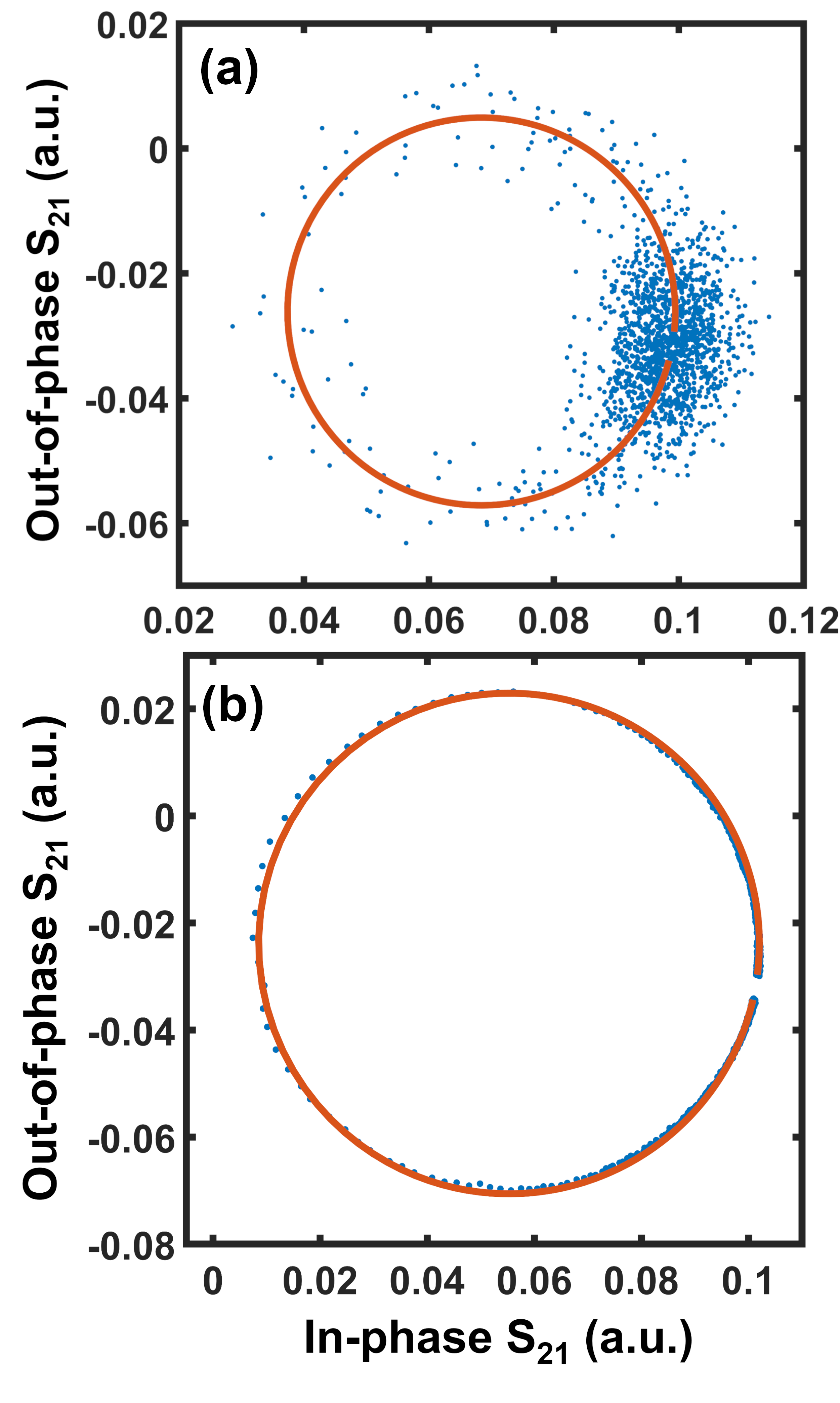}
    \caption{Representative $S_{21}$ data and fits (red curve) of a Ta resonator with $N_{AB}$ = 0 for (a) $\langle n_p \rangle \sim 1$ and (b) $\langle n_p \rangle \sim 10^5$. \label{FigS4}}
\end{figure*}

The measured in-phase and out-of-phase $S_{21}$ versus frequency $f$ were simultaneously fitted using the diameter correction method \cite{2012Khalil}
\begin{equation}
    S_{21}(f) = (S_{x}+iS_{y})\left(1-\frac{Q_L/\left|\hat{Q_c}\right| \cos\phi}{1+i\frac{Q_L}{ff_r}\left(f^2-f_r^2\right)}e^{i\phi}\right).
    \label{eq:S21}
\end{equation}
Here, $S_{x}$ and $S_{y}$ are associated with the off resonance background transmitted signal through the system, $f_r$ is the resonance frequency, $Q_L$ is the loaded quality factor, and $\hat{Q_c}$ is the complex coupling quality factor with phase $\phi$. The internal quality factor $Q_i$ is subsequently extracted by
\begin{equation}
    \frac{1}{Q_i} = \frac{1}{Q_L} - \frac{1}{\left|\hat{Q_c}\right|}.
    \label{eq:Q_relation}
\end{equation}
Fig. \ref{FigS4} shows two representing fits (red curve) with an average stored photon number $\langle n_p \rangle \sim 1$ (Fig. \ref{FigS4}(a)) and $\langle n_p \rangle \sim 10^5$ (Fig. \ref{FigS4}(b)), for the Ta resonator without air bridges. To estimate the power at the device and thus $\langle n_p \rangle$, a nominal attenuation of 70 dB  was used.

\section{Voxel size}
\begin{figure*}
    \centering
    \includegraphics[width=.5\linewidth]{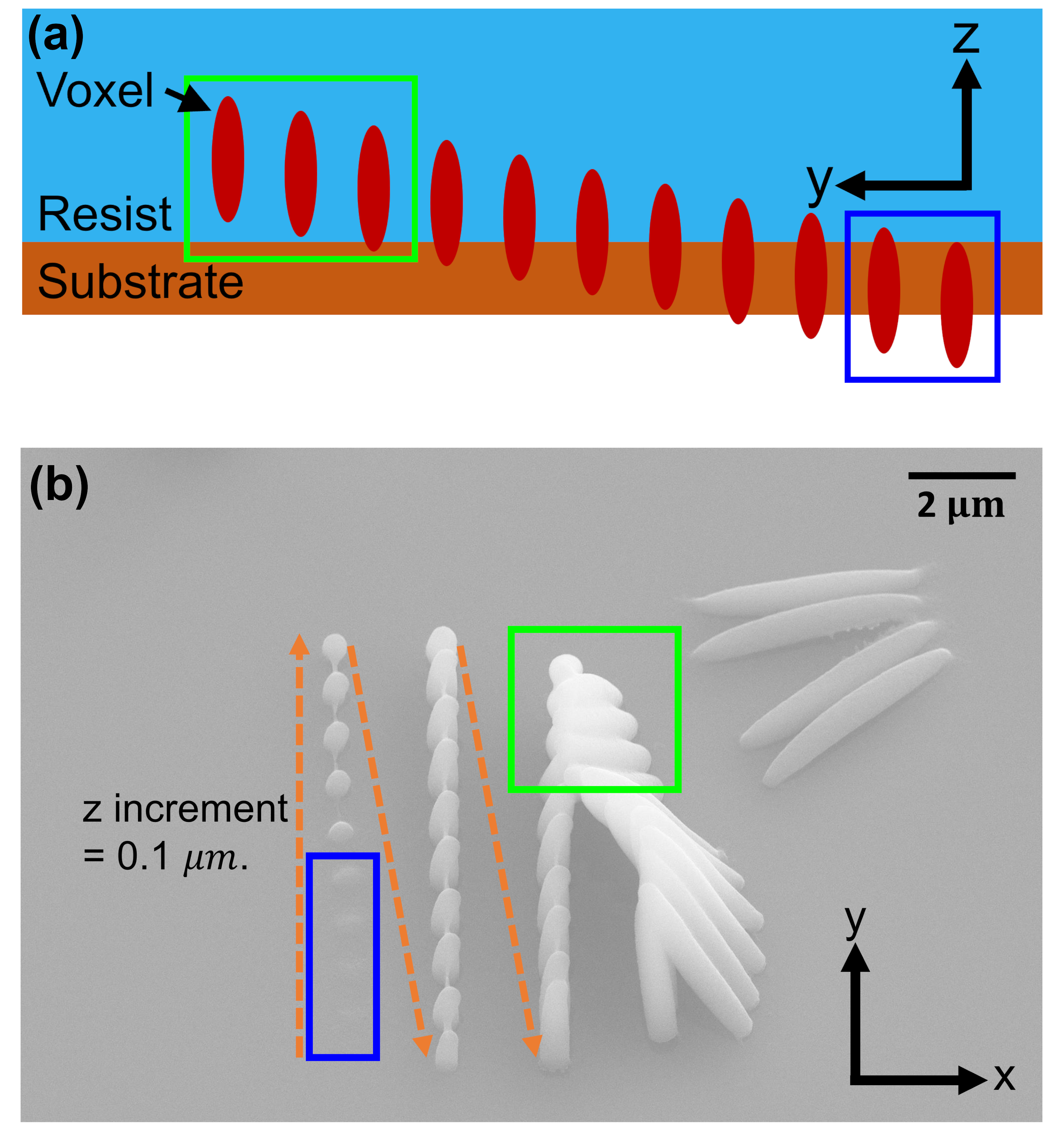}
    \caption{Experimental determination of the voxel size of the AZ 15nXT photoresist. (a) A series of voxels are created by moving the laser in the z direction with an increment of 0.1 $\mu$m. (b) SEM image of the voxels. The z position of the voxels increases in steps of 0.1 $\mu$m along the dashed lines. Blue box: light is focused inside the substrate and the voxels are not formed. Green box: voxels collapse due to lack of support. \label{FigS1}}
\end{figure*}

The voxel size of the AZ 15nXT photoresist was experimentally determined. To create a single voxel, the laser was focused at a fixed point in the resist/substrate and the shutter was opened for 10 ms, allowing the resist to polymerize. By stepping the focal point at different heights (z direction) starting from below the substrate-resist interface  in a sequential manner, a series of voxels were created (see Fig. \ref{FigS1}(a)). Fig. \ref{FigS1}(b) shows a SEM image of a series of voxels with their z position increasing, as indicated by the dashed lines, in steps of 0.1 $\mu$m. 

Two regions are of particular interest in a voxel test. When the light is nominally focused inside the substrate, the resist is not fully exposed to form a voxel (Fig. \ref{FigS1} blue box). On the other hand, when the voxel is focused inside the resist and above the substrate, the corresponding structure will collapse due to lack of support (Fig. \ref{FigS1} green box).
From the SEM image,  an upper bound for the lateral (xy direction) size of the voxel is estimated  to be 0.5 $\mu$m, and an upper bound for the axial (z direction) size is estimated to be 3 $\mu$m.

\section{Electromagnetic Simulations} \label{EM_sim}
Electromagnetic microwave simulations using Ansys' high frequency simulation software (HFSS) were performed of the devices to estimate the participation of inductive energy stored in the AB ($p$) and the loss at higher temperatures. To simulate these quantities we are interested in the magnetic field $H$ at the  surfaces of the different conductors.

\begin{figure*}
    \centering
    \includegraphics[width=.5\linewidth]{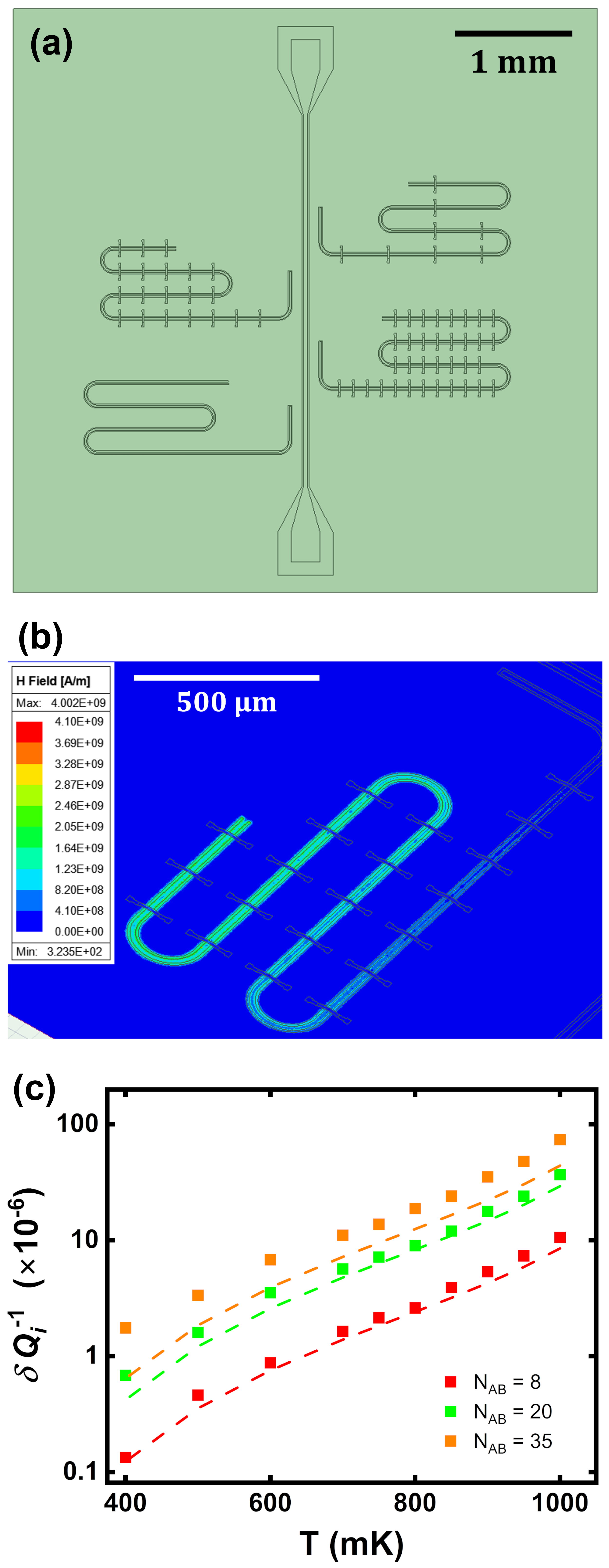}
    \caption{(a) CAD rendering of a chip containing four resonators with $N_{AB}$ = 0, 8, 20, 35, respectively, for HFSS simulation. (b) $H$ field distribution for a resonator with $N_{AB}$ = 20. (c) $\delta Q_i^{-1}$ and simulated loss (dashed line) as a function of T for various $N_{AB}$. \label{FigS3}}
\end{figure*}

Fig. \ref{FigS3}(a) shows the CAD rendering of the resonator/AB chip used for HFSS simulation. To reduce computational resources, the ground plane, coplanar waveguides, and air bridges are assumed to be a perfect electric conductor. The sapphire substrate had a dimension of 5 mm $\times$ 5 mm $\times$ 0.5 mm and a dielectric constant $\epsilon_r = 10$. The simulation space is defined as a 5 mm $\times$ 5 mm $\times$ 5 mm cube that contains the whole chip. Fig. \ref{FigS3}(b) shows the simulated ac $H$ field distribution excited on resonance of the resonator with 20 AB. 

The parameter $p$ is given by
\begin{equation}
    p = \frac{\int_{\mathrm{AB}}|H|^2 \, ds}{\int_{\mathrm{chip}}|H|^2 \, ds + \int_{\mathrm{AB}}|H|^2 \, ds}.
\end{equation}
Here the limits of the surface integral ``chip'' is evaluated at the metal interfaces of the chip (\textit{e.g.} ground plane and coplanar traces) but excludes the AB, while the limits of the integral ``AB'' is only of the surface of the AB. Performing the simulations and integrals we find $p_{N_{AB} = 8}=0.6\%$, $p_{N_{AB} = 20}=2.0\%$, and $p_{N_{AB} = 35}=3.4\%$; $p$ values that are 20 to 30\% smaller than the parameters measured from internal losses at elevated temperatures in the text. 

To simulate the conducting losses associated with quasiparticles in the Al AB  at temperatures $T>400$ mK, the dissipated power from induced currents in the AB were calculated using\cite{2005Pozar}
\begin{equation}
    P_{s} = \frac{R_s}{2}\int_{\mathrm{AB}}|H|^2 \, ds.
\end{equation}
Here we evaluate the surface resistance $R_s$ of the Al AB in the bulk limit using \cite{1977Broom}
\begin{equation}
    R_s = 2\pi f_r \mu_0 \lambda \frac{\sigma_1}{2\sigma_2}
\end{equation}
with $\lambda$, $\sigma_1$, and $i\sigma_2$ defined in Eq. \ref{eq:MB} of the main text. \par
The limiting quality factor $Q_{s}$ is then given by \cite{2023Huang} 
\begin{equation}
    Q_{s}^{-1} = \frac{R_s\int_{\mathrm{AB}}|H|^2 \, ds}{2\pi f_r \mu_0 \int_{\mathrm{simulation\,space}} |H|^2 \, dv}.
    \label{eq:Qs}
\end{equation}
This simulated loss from thermally generated quasiparticles in the Al AB can then be compared to the measured loss at temperatures between 400 mK and 1.2 K. To do this comparison, we  subtract off the losses from the bare Ta resonator using the 0 AB resonator data
\begin{equation}
    \delta Q_i^{-1} = Q^{-1}_{i, \, N_{AB} \,=\, 8,\, 20,\, 35}(T) - Q^{-1}_{i, \, N_{AB} \,=\, 0} (T).
\end{equation}
Fig. \ref{FigS3}(c) shows $\delta Q_i^{-1}$  as a function of $T$ for the three resonators with $N_{AB}\neq0$. Note that as $N_{AB}$ increases, the measured losses $\delta Q_{i}^{-1}$ also increases due to more energy stored in the Al AB. The simulated losses (dashed curves in Fig. \ref{FigS3}(c)), while in good agreement to the loss data, appear to undercount $p$ in the AB similar to our $p$ estimates.

\section{Resonance Frequency Shift at Elevated Temperatures}
\begin{figure*}
    \centering
    \includegraphics[width=.5\linewidth]{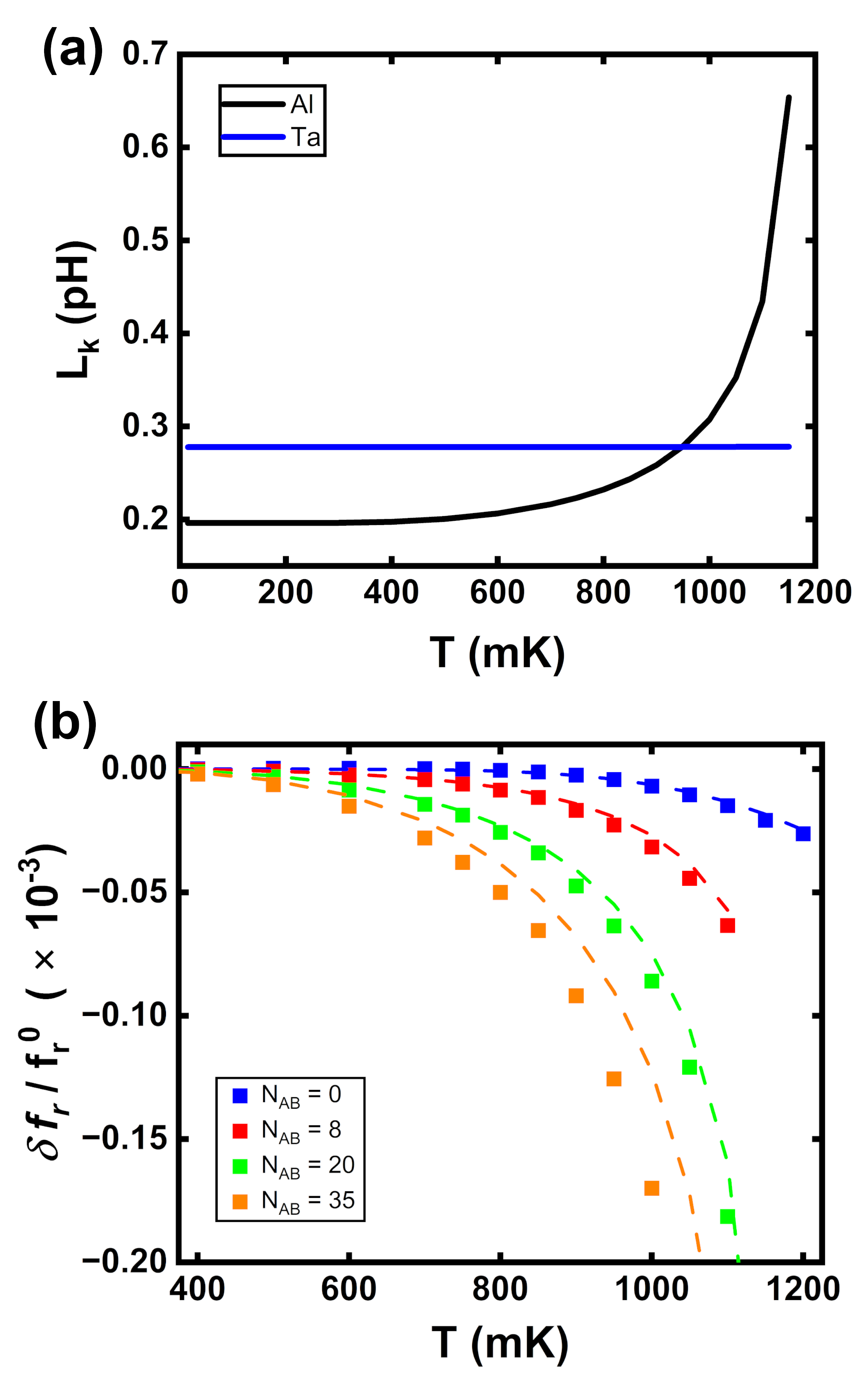}
    \caption{(a) The kinetic inductance ($L_k$) as a function of T for Al (black) and Ta (blue). (b) Fractional frequency shift $\delta f_r / f_r^0$ as a function of temperature at $\langle n_p \rangle\simeq 1$. With more AB added, the resonance frequencies shift more at high temperatures. Dashed line shows the theoretically calculation using Eq. \ref{eq:dfr}. \label{FigS5}}
\end{figure*}

The shifts in resonant frequencies at elevated temperatures from $T = 400$ mK to above 1 K were measured. The resonance frequency $f_r$ of a quarter-wave CPW resonator is given by
\begin{equation}
    f_r = \frac{1}{4l\sqrt{(L_g+L_k)C}}
    \label{eq:fr}
\end{equation}
where $L_k$ ($L_g$) is the kinetic (geometric) inductance and $l$ is the length of the CPW resonator. At elevated temperatures, $L_k=\mu_0\lambda$ increases as more quasiparticles are thermally excited. Fig. \ref{FigS5}(a) shows the theoretically calculated $L_k$ as a function of $T$ for Al (black curve) and Ta (blue curve) in the bulk homogeneous limit. The kinetic inductance of Al, $L_k^{Al}$, increases and surpasses $L_k^{Ta}$ as the temperature approaches 1.2 K, the superconducting transition temperature of Al. \par
Fig. \ref{FigS5}(b) shows the measured fractional frequency shift $\delta f_r / f_r^0 \equiv (f_r(T) - f_r^0) / f_r^0$ as a function of $T$ at $\langle n_p \rangle\simeq 1$ ($f_r^0$ denotes the resonance frequency at $T = 10$ mK). A clear dependence of $\delta f_r / f_r^0$ on $N_{AB}$ is observed: with more AB added, $\delta f_r / f_r^0$ changes more at high temperatures. Using Eq. \ref{eq:fr},  the fractional frequency shift is calculated to be
\begin{equation}
    \delta f_r / f_r^0 = \frac{\sqrt{L_g+L_k^{Ta}(T = 10 \, \mathrm{mK})}}{\sqrt{L_g+L_k^{Ta}(T)+p_{N_{AB}}L_k^{Al}(T)}} -1 
    \label{eq:dfr}
\end{equation}
where $p_{N_{AB}}$ is the participation of AB estimated in Sec. \ref{EM_sim}. Using $L_g=16\,\mathrm{pH}$,  the calculated values of $\delta f_r / f_r^0$ are shown in Fig. \ref{FigS5}(b) (dashed line) for the four resonators.

\section*{References}
\vspace{-12pt}
\bibliography{ref}

